\documentclass[a4paper,11pt]{article}
\usepackage{graphicx}
\usepackage{psfrag}
\usepackage{amsmath}
\usepackage{amssymb}
\usepackage{feynarts}
\usepackage{cite}
\usepackage{mcite}

\begin{document}
\newcommand{\beq}{\begin{eqnarray}}
\newcommand{\eeq}{\end{eqnarray}}
\newcommand{\non}{\nonumber}
\newcommand{\mc}{\mathcal}
\newcommand{\ra}{\rightarrow}

\title{\ \\ \ \\ 
\LARGE{\bf{Radiative Corrections to $H^0\to WW/ZZ$ in the MSSM}}}
\author{\Large{Wolfgang Hollik${}^{\,a}$ and {Jian-Hui Zhang${}^{\,a,b}$}}}
\date{${}^{a}$\it{Max-Planck-Institut f\"ur Physik, \\
F\"ohringer Ring 6, 80805 M\"unchen, Germany}
\vskip .3em
      ${}^{b}$Center for High Energy Physics, \\
      Peking University, Beijing, China}

\maketitle

\begin{abstract}
The electroweak $\mc O(\alpha)$ radiative corrections to the decay of the heavy CP-even MSSM Higgs boson to weak gauge bosons are presented. Due to the suppression of the tree-level $H^0WW/H^0ZZ$ coupling, the electroweak contributions to the partial decay width are significant. Although the effective Born decay width can be rather small for certain $M_{A^0}$ values (especially for large $\tan\beta$), the corrected partial widths at different values of $M_{A^0}$ are of comparable size.
\end{abstract}

\section{Introduction}
One of the major tasks of the Large Hadron Collider (LHC) at CERN is to explore the mechanism responsible for the  electroweak symmetry breaking (EWSB). In the standard model (SM), the EWSB is realized by the Higgs mechanism, which predicts the existence of a physical scalar boson, the Higgs boson. The mass of the Higgs boson cannot be predicted by the SM. Experimental searches at LEP~\cite{Barate:2003sz} and Tevatron~\cite{Aaltonen:2011gs} have excluded a Higgs mass $M_H< 114.4$\,GeV and $158\,\mbox{GeV}<M_H<173\,\mbox{GeV}$, both at $95\%$ confidence level (C.L.). In addition, electroweak precision analysis favors a relatively light SM Higgs boson with a mass $M_H<158$\,GeV (or $M_H<185$\,GeV if the LEP2 direct search limit is included)~\cite{LEP:2008wb}.

In the minimal supersymmetric standard model (MSSM), which is the most economic extension of the SM that incorporates supersymmetry, two complex Higgs doublets are required for consistency. After the EWSB, three of the eight degrees of freedom are absorbed by the weak gauge bosons and become their longitudinal components, leading to five physical Higgs bosons, $h^0$ and $H^0$ (CP-even), $A^0$ (CP-odd) and $H^\pm$ (charged). At tree-level the Higgs sector of MSSM can be described by two parameters, which can be chosen as $M_{A^0}$ and $\tan\beta$, where $M_{A^0}$ is the mass of the CP-odd Higgs boson and $\tan\beta$ is the ratio of the vacuum expectation values of the two Higgs doublets. The tree-level mass of the light CP-even Higgs boson is bounded from above by the mass of $Z$ boson as a consequence of supersymmetry. Dependence of the Higgs sector on parameters of other sectors enters via radiative corrections. By including radiative corrections (up to two-loop order), the upper mass bound of the light CP-even Higgs boson is shifted to $\sim135$\,GeV~\cite{Heinemeyer:1998jw,*Heinemeyer:1998kz,*Heinemeyer:1998np,Degrassi:2002fi,Allanach:2004rh}.

It is interesting to investigate the behavior of the MSSM Higgs sector in the decoupling limit $M_{A^0}\gg M_Z$~\cite{Gunion:2002zf,*Carena:2002es}. In this limit, the heavy CP-even Higgs boson $H^0$ and the charged Higgs bosons $H^\pm$ are nearly degenerate in mass with $A^0$, and the coupling of the light CP-even Higgs boson to SM fermions and gauge bosons resembles the corresponding coupling for the SM Higgs boson. If only one light Higgs boson is discovered, it might be compatible both with the SM and with the MSSM. If there are extra heavy Higgs bosons observed, they would indicate the incompleteness of the SM. Investigating the decay properties of such heavy Higgs bosons can then help disentangle supersymmetry from other potential extensions of the SM with an extended Higgs sector.

In the MSSM, the tree-level coupling of the heavy CP-even Higgs boson to gauge bosons is suppressed by a factor of $\cos(\beta-\alpha)$, compared to the corresponding coupling for the SM Higgs boson, where $\alpha$ is the angle that diagonalizes the CP-even Higgs sector at tree-level. This suppression can be rather strong for large values of $M_{A^0}$, hence it is of particular interest to investigate the impact of radiative corrections in such situations.

In this paper we consider the electroweak radiative corrections to the decay of the heavy CP-even Higgs boson to weak gauge bosons. Electroweak corrections can induce important modifications to the tree-level $H^0WW/H^0ZZ$ coupling. One potential source of large corrections is the contribution of loops involving fermions and sfermions, and in particular of loops involving the third generation fermions and sfermions, since they contain potentially large Yukawa couplings. The Higgs propagator corrections can give rise to significant contributions as well. In this work we concentrate on CP-conserving MSSM with real parameters, therefore the heavy CP-even Higgs boson can only mix with the light CP-even one.

The outline of this paper is as follows. In section~\ref{HVVvert} we discuss the structure of the $H^0WW/H^0ZZ$ vertex correction. In section~\ref{HVVonshell} the decay amplitudes of $H^0$ to $WW/ZZ$ are given. Section~\ref{numer} is devoted for numerical discussions. We draw our conclusions in section~\ref{conclusion}.

\section{Corrections to the $H^0WW/H^0ZZ$ vertices}
\label{HVVvert}
\subsection{Correction to the $H^0WW$ vertex}
The tree-level $H^0WW$ coupling is given by
\beq\label{HWWtreecoupl}
V^{\mu\nu}_{H^0,0}=\frac{eM_W}{s_W}\cos(\beta-\alpha)g^{\mu\nu}\equiv V_{H^0}\cos(\beta-\alpha)g^{\mu\nu}\ ,
\eeq
where $V_{H^0}$ denotes the coupling of the SM Higgs boson to $W$ bosons. When the mass of the CP-odd Higgs boson $M_{A^0}$ becomes large, the angle $\beta-\alpha\to\pi/2$ and the factor $\cos(\beta-\alpha)$ approaches 0 as $\mc O(M_Z^2|\sin4\beta|/M_{A^0}^2)$, thus the tree-level coupling is strongly suppressed.

\begin{figure}[htbp]
\centering
\includegraphics[width=0.8\textwidth]{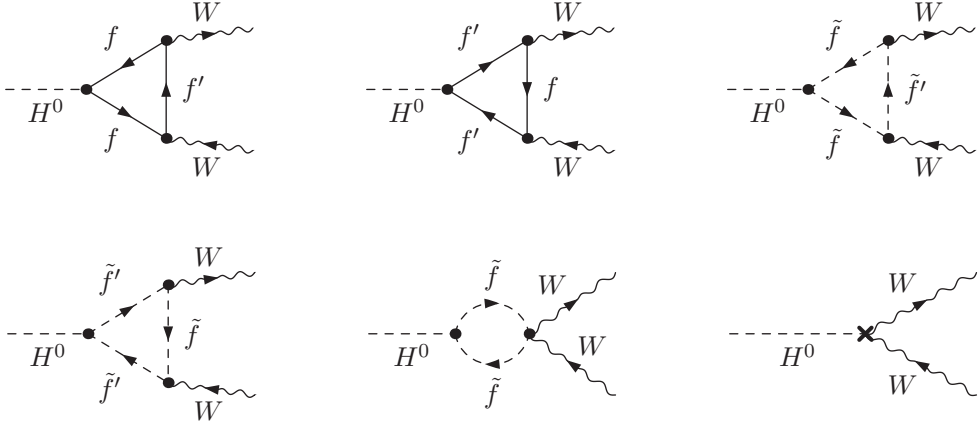}
\vskip -0.6cm
\caption{One-loop vertex diagrams and counter term diagram for $H^0WW$.}
\label{HWWvc}
\end{figure}

The one-loop corrected $H^0WW$ vertex possesses the following structure~\cite{Kniehl:1990mq,*Kniehl:1991xe}
\begin{align}\label{1loopHWW}
V^{\mu\nu}_{H^0}&=V^{\mu\nu}_{H^0,0}+V^{\mu\nu}_{H^0,1}\non\\
&=V^{\mu\nu}_{H^0,0}+V_{H^0}(Ak_2^\mu k_2^\nu+Bk_3^\mu k_3^\nu+Ck_2^\mu k_3^\nu+Dk_3^\mu k_2^\nu+Eg^{\mu\nu}-iF\varepsilon^{\mu\nu\rho\sigma}k_{2\rho}k_{3\sigma})\ ,
\end{align}
where $k_2$, $k_3$ denote the momenta of the two gauge bosons, $\varepsilon^{\mu\nu\rho\sigma}$ is totally antisymmetric with $\varepsilon^{0123}=1$. If the gauge bosons are on-shell, then only $D$ and $E$ terms will contribute. As mentioned in the introduction, the loop contributions from the fermionic and sfermionic sector, and in particular from the third generation fermions and sfermions are expected to be sizable. In Fig.~\ref{HWWvc} we show as examples the Feynman diagrams involving fermion and sfermion loops. The packages FeynArts~\cite{Kublbeck:1990xc,*Hahn:2000kx,*Kublbeck:1992mt}. FormCalc~\cite{Hahn:2001rv,*Hahn:2006qw,*Hahn:2007px} and LoopTools~\cite{Hahn:1999wr} are used throughout the computation.

The contribution of the vertex counter term diagram in Fig.~\ref{HWWvc} can be written as (see {\it e.g.}~\cite{Hahn:2002gm})
\begin{align}\label{HWWct}
V_{H^0,CT}^{\mu\nu}&=V_{H^0,0}^{\mu\nu}\Big[\delta Z_e+\delta Z_W+\frac{1}{2}\frac{\delta M_W^2}{M_W^2}+\frac{\delta s_W}{s_W}+\frac{\sin(\beta-\alpha)}{\cos(\beta-\alpha)}(-\cos^2\beta\,\delta\tan\beta+\frac{1}{2}\delta Z_{h^0H^0})\non\\
&+\frac{1}{2}\delta Z_{H^0H^0}
\Big]\ ,
\end{align}
where $\delta Z_e$ is the charge renormalization constant, $\delta Z_W$ is the field renormalization constant for the $W$ boson, and $\delta M_W^2$ the mass counter term for the $W$ boson. $\delta s_W$ is the renormalization constant for the weak mixing angle. $\delta Z_{h^0H^0}$ and $\delta Z_{H^0H^0}$ are the Higgs field renormalization constants in the mass eigenstate basis. The counter term for $\tan\beta$ is introduced via $\tan\beta\to \tan\beta+\delta\tan\beta$. The vertex counter term is proportional to the tree-level coupling, and thus contributes to the $E$-term in Eq.~(\ref{1loopHWW}).

To determine the Higgs field renormalization constants, we choose the $\overline{DR}$ scheme~\cite{Frank:2006yh}. In this scheme, the Higgs field renormalization constants are given by
\begin{align}\label{Hfieldrenconst}
\delta Z_{h^0h^0}&=-[\mbox{Re}\,\Sigma'_{h^0h^0}(m_{h^0}^2)]^{\mbox{\small{div}}}\ ,\non\\
\delta Z_{H^0H^0}&=-[\mbox{Re}\,\Sigma'_{H^0H^0}(m_{H^0}^2)]^{\mbox{\small{div}}}\ ,\non\\
\delta Z_{h^0H^0}&=\delta Z_{H^0h^0}=\frac{\sin\alpha\cos\alpha}{\cos 2\alpha}(\delta Z_{h^0h^0}-\delta Z_{H^0H^0})\ ,
\end{align}
where $m_{h^0}$, $m_{H^0}$ denote the tree-level masses of the two CP-even Higgs bosons, "div" means we only take into account the divergent parts of the renormalization constants. The counter term $\delta\tan\beta$ is fixed by
\beq
\frac{\delta\tan\beta}{\tan\beta}=\left(\frac{\delta\tan\beta}{\tan\beta}\right)^{\overline{DR}}=\frac{1}{2\cos 2\alpha}(\delta Z_{h^0h^0}-\delta Z_{H^0H^0})\ .
\eeq
For the renormalization scale, we choose $\mu^{\overline{DR}}=m_t$. The remaining counter terms in Eq.~(\ref{HWWct}) are fixed in the on-shell scheme as follows~\cite{Denner:1991kt}
\begin{align}\label{OSrenconst}
\delta Z_e&=\frac{1}{2}{\Sigma'}_\gamma(0)-\frac{s_W}{c_W}\frac{\Sigma_{\gamma Z}^T(0)}{M_Z^2}\ ,\non\\
\delta Z_W&=-\mbox{Re}\,{\Sigma'}_W^{T}(M_W^2)\ ,\non\\
\delta M_W^2&=\mbox{Re}\,\Sigma_W^T(M_W^2)\ ,\non\\
\delta M_Z^2&=\mbox{Re}\,\Sigma_Z^T(M_Z^2)\ ,\non\\
\frac{\delta s_W}{s_W}&=\frac{1}{2}\frac{c_W^2}{s_W^2}(\frac{\delta M_Z^2}{M_Z^2}-\frac{\delta M_W^2}{M_W^2})\ ,
\end{align}
where the prime indicates the derivative and the superscript $T$ denotes the transverse part of the corresponding self energy. 

Due to the mixing between the two CP-even Higgs bosons beyond tree-level, when evaluating the radiative corrections to the decay $H^0\to WW$ we need to take into account the $h^0WW$ vertex correction as well. The tree-level $h^0WW$ coupling is simply obtained by replacing $\cos(\beta-\alpha)$ in Eq.~(\ref{HWWtreecoupl}) with $\sin(\beta-\alpha)$, and the corresponding counter term contribution reads
\begin{align}
V_{h^0,CT}^{\mu\nu}&=V_{H^0}\sin(\beta-\alpha)g^{\mu\nu}\Big[\delta Z_e+\delta Z_W+\frac{1}{2}\frac{\delta M_W^2}{M_W^2}+\frac{\delta s_W}{s_W}+\frac{\cos(\beta-\alpha)}{\sin(\beta-\alpha)}(\cos^2\beta\,\delta\tan\beta\non\\
&+\frac{1}{2}\delta Z_{H^0h^0})+\frac{1}{2}\delta Z_{h^0h^0}
\Big]\ ,
\end{align}
where the field renormalization constants $\delta Z_{h^0 h^0}$ and $\delta Z_{H^0 h^0}$ are given in Eq.~(\ref{Hfieldrenconst}).

\subsection{Correction to the $H^0ZZ$ vertex}
The tree-level $H^0ZZ$ coupling has the same structure as the tree-level $H^0WW$ coupling
\beq\label{HZZtreecoupl}
V_{H^0,0}^{'\mu\nu}=\frac{eM_W}{c_W^2 s_W}\cos(\beta-\alpha)g^{\mu\nu}\equiv V_{H^0}'\cos(\beta-\alpha)g^{\mu\nu}\ ,
\eeq
it also differs from the SM counterpart by a factor of $\cos(\beta-\alpha)$ and gets strongly suppressed in the decoupling limit $M_{A^0}\gg M_Z$.

The corrected vertex has the same tensor structure as in Eq.~(\ref{1loopHWW}), with $V_{H^0}$ replaced by $V_{H^0}'$. The corresponding counter term contributions are given by
\begin{align}\label{HZZct}
V_{H^0,CT}^{'\mu\nu}&=V_{H^0,0}^{'\mu\nu}\Big[\delta Z_e+\delta Z_Z+\frac{1}{2}\frac{\delta M_W^2}{M_W^2}-\frac{\delta s_W}{s_W}(2\frac{s_W^2}{c_W^2}-1)+\frac{\sin(\beta-\alpha)}{\cos(\beta-\alpha)}(-\cos^2\beta\,\delta\tan\beta\non\\
&+\frac{1}{2}\delta Z_{h^0H^0})+\frac{1}{2}\delta Z_{H^0H^0}
\Big]\ ,
\end{align}
and 
\begin{align}
V_{h^0,CT}^{'\mu\nu}&=V_{H^0}'\sin(\beta-\alpha)g^{\mu\nu}\Big[\delta Z_e+\delta Z_Z+\frac{1}{2}\frac{\delta M_W^2}{M_W^2}-\frac{\delta s_W}{s_W}(2\frac{s_W^2}{c_W^2}-1)\non\\
&+\frac{\cos(\beta-\alpha)}{\sin(\beta-\alpha)}(\cos^2\beta\,\delta\tan\beta+\frac{1}{2}\delta Z_{H^0h^0})+\frac{1}{2}\delta Z_{h^0h^0}
\Big]\ ,
\end{align}
with $\delta Z_Z=-\mbox{Re}\,{\Sigma'}_Z^{T}(M_Z^2)$.

\section{Decay amplitudes of $H^0\to WW/ZZ$}
\label{HVVonshell}
The decay amplitude can be obtained from the coupling vertex in the previous section by putting all external momenta on-shell and multiplying with the polarization vectors for the external gauge bosons. Beyond the lowest order, the CP-even Higgs propagator matrix receives important radiative corrections, leading to finite wave function normalization factors for the external Higgs boson. These wave function normalization factors can be incorporated by using the following effective Born amplitude
\begin{align}\label{oseffbornamp}
\mc M^0_{\mbox{eff}}&=\sqrt{Z_{H^0}}(\mc M_{H^0}^0+Z_{H^0h^0}\mc M_{h^0}^0)\non\\
&=\sqrt{Z_{H^0}}\mc M_{H^0}^0(1+\tan(\beta-\alpha)Z_{H^0h^0})\ ,
\end{align}
where $\mc M_{H^0}^0$ and $\mc M_{h^0}^0$ denote the tree-level decay amplitude for $H^0$ and $h^0$, respectively. For illustrating purposes, we will present in the next section both the tree-level results obtained from $\mc M_{H^0}^0$ and the effective tree-level results obtained from $\mc M^0_{\mbox{eff}}$.  The wave function normalization factors $Z_{H^0}$ and $Z_{H^0h^0}$ in Eq.~(\ref{oseffbornamp}) can be determined from the renormalized self energies of Higgs bosons as
\begin{align}
Z_{H^0}&=\frac{1}{1+\mbox{Re}\hat\Sigma_{H^0}'(k^2)-\mbox{Re}\left(\frac{\hat\Sigma_{h^0H^0}^2(k^2)}{k^2-m_{h^0}^2+\hat\Sigma_{h^0}(k^2)}\right)'}\Big|_{k^2=M^2_{H^0}}\ ,\non\\
Z_{H^0h^0}&=-\frac{\hat\Sigma_{h^0H^0}(M^2_{H^0})}{M^2_{H^0}-m_{h^0}^2+\hat\Sigma_{h^0}(M^2_{H^0})}\ ,
\end{align}
where $M_{H^0}$ is the physical mass of $H^0$. In this work the physical masses of Higgs bosons and the finite wave function normalization factors are computed with the program package FeynHiggs~\cite{Heinemeyer:1998yj,*Hahn:2005cu,*Hahn:2006np}, in which the dominant two-loop corrections to the Higgs boson self energies are also taken into account. For the decay to $W$ bosons, $\mc M_{H^0}^0$ is given by
\beq
\mc M_{H^0}^0=V_{H^0,0}^{\mu\nu}\epsilon_\mu\epsilon_\nu\ ,
\eeq
where $V_{H^0,0}^{\mu\nu}$ is defined in Eq.~(\ref{HWWtreecoupl}) and $\epsilon_{\mu,\nu}$ are the polarization vectors of the external $W$ bosons. The amplitude for the decay of $H^0$ to $Z$ bosons can be obtained analogously. 

At one-loop level, one can also write an effective amplitude 
\beq
\mc M^1_{\mbox{eff}}=\sqrt{Z_{H^0}}(\mc M_{H^0}^1+Z_{H^0h^0}\mc M_{h^0}^1)\ .
\eeq
In the following the results obtained from this effective amplitude will be denoted as effective one-loop results. In obtaining such results we also include the square of the one-loop amplitude, since the tree-level coupling can be suppressed so that the square of the one-loop amplitude can become comparable to the tree-level result. For the decay $H^0\to WW$, we compute the complete $\mc O(\alpha)$ contributions, and the effective one-loop contributions from the fermionic and sfermionic loops, which do not involve infrared divergences. For the decay $H^0\to ZZ$, we compute also the complete effective one-loop contribution as the complete one-loop amplitude is infrared finite in this case.

\section{Numerical discussions}
\label{numer}
For the numerical evaluation, we choose the benchmark scenarios suggested in~\cite{Carena:1999xa,*Carena:2002qg}, where the two parameters that govern the tree-level Higgs sector, $M_{A^0}$ and $\tan\beta$, are kept as free parameters. The SM parameters used in the numerical analysis are the same as in ref.~\cite{Hollik:2010ji}. Throughout the parameter scan, the experimental mass exclusion limits from direct search of supersymmetric particles and the upper bound on the SUSY corrections to the electroweak $\rho$ parameter~\cite{Yao:2006px} have been taken into account (except for the tree-level result). In the parameter region $50\,\mbox{GeV}<M_{A^0}<1\,\mbox{TeV}$ and $1<\tan\beta<50$, the gluophobic scenario has been ruled out by the bound derived from the $\mbox{BR}(B\to X_s\gamma)$ prediction~\cite{Brein:2007da}, therefore we will not discuss this scenario here. The investigated scenarios are listed as follows:

\noindent 1. The $m_h^{\mbox{\small{max}}}$ scenario \\
The parameters in this scenario are given by
\begin{align}
M_{\mbox{\tiny{SUSY}}}&=1\,\mbox{TeV}\ , & \mu&=200\,\mbox{GeV}\ , & M_2&=200\,\mbox{GeV} \ ,\non\\
X_t&=2M_{\mbox{\tiny{SUSY}}}\ , & A_b&=A_t=A_{\tau}\ , & m_{\tilde g}&=0.8M_{\mbox{\tiny{SUSY}}}\ ,
\end{align}
where $M_{\mbox{\tiny{SUSY}}}$ is the soft SUSY-breaking parameter, $\mu$ is the supersymmetric Higgs mass parameter,  $M_2$ denotes the $SU(2)$ gaugino mass, $X_t$ the mixing parameter of the top squark sector, $A_{b, t, \tau}$ the trilinear couplings for the third generation squark and slepton and $m_{\tilde g}$ the gluino mass. This scenario yields a maximal value of the lightest CP-even Higgs boson mass for given $M_{A^0}$ and $\tan\beta$.

\noindent 2. The no-mixing scenario \\
The only difference of this scenario from the $m_h^{\mbox{\small{max}}}$ scenario is the vanishing mixing in the top squark sector and a higher value of $M_{\mbox{\tiny{SUSY}}}$, where the latter is chosen to avoid the exclusion bounds from the LEP Higgs searches~\cite{Barate:2003sz,Schael:2006cr}. The parameters in this scenario read
\begin{align}
M_{\mbox{\tiny{SUSY}}}&=2\,\mbox{TeV}\ , & \mu&=200\,\mbox{GeV}\ , & M_2&=200\,\mbox{GeV} \ ,\non\\
X_t&=0\ , & A_b&=A_t=A_{\tau}\ , & m_{\tilde g}&=0.8M_{\mbox{\tiny{SUSY}}}\ .
\end{align}

\noindent 3. The small-$\alpha_{\mbox{\small{eff}}}$ scenario \\
In this scenario a suppression of the $h^0b\bar b$ coupling can occur. The parameters are given by
\begin{align}
M_{\mbox{\tiny{SUSY}}}&=800\,\mbox{GeV}\ , & \mu&=2.5M_{\mbox{\tiny{SUSY}}}\ , & M_2&=500\,\mbox{GeV} \ ,\non\\
X_t&=-1100\,\mbox{GeV}\ , & A_b&=A_t=A_{\tau}\ , & m_{\tilde g}&=500\,\mbox{GeV}\ .
\end{align}

\begin{figure}[htbp]
\includegraphics[width=0.47\textwidth]{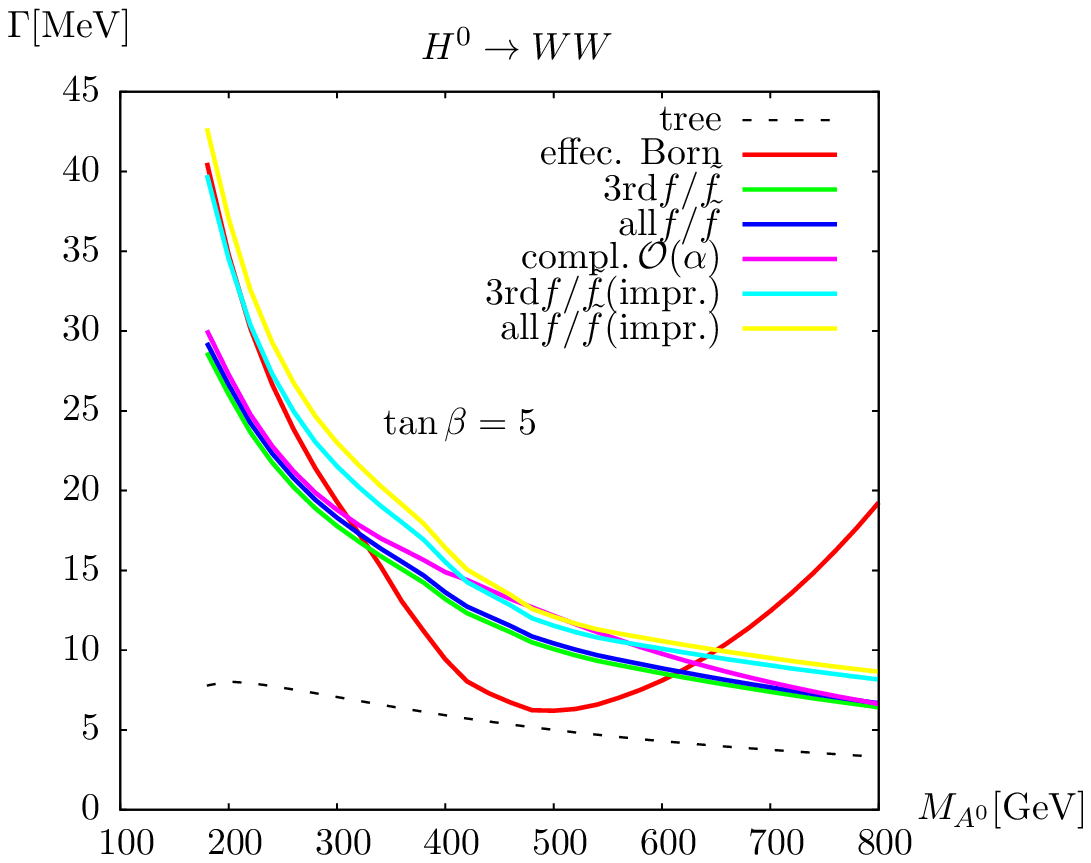}
\hspace{2em}
\includegraphics[width=0.47\textwidth]{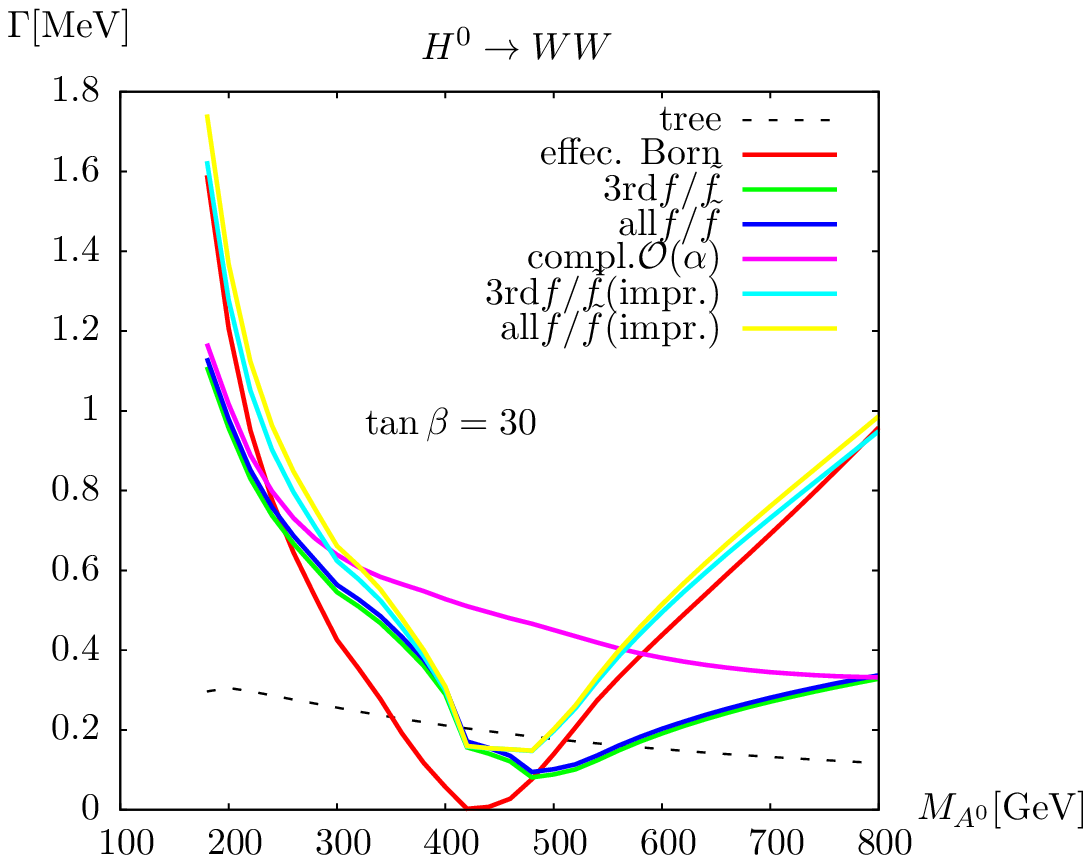}
\caption{The lowest order and the corrected partial decay widths for $H^0\to WW$ as a function of $M_{A^0}$ in the $m_h^{\mbox{\small{max}}}\;\mbox{scenario}$ for $\tan\beta=5, 30$, where "effec. Born" denotes the effective Born result, "3rd (all) $f/\tilde f$" denote the results including contributions from the third (all) generation fermions and sfermions, "compl. $\mc O(\alpha)$" means the result including the complete $\mc O(\alpha)$ contribution, "impr." denotes the effective one-loop result. }
\label{HWWwidthmaxcompTB}
\end{figure}

\begin{figure}[htbp]
\includegraphics[width=0.47\textwidth]{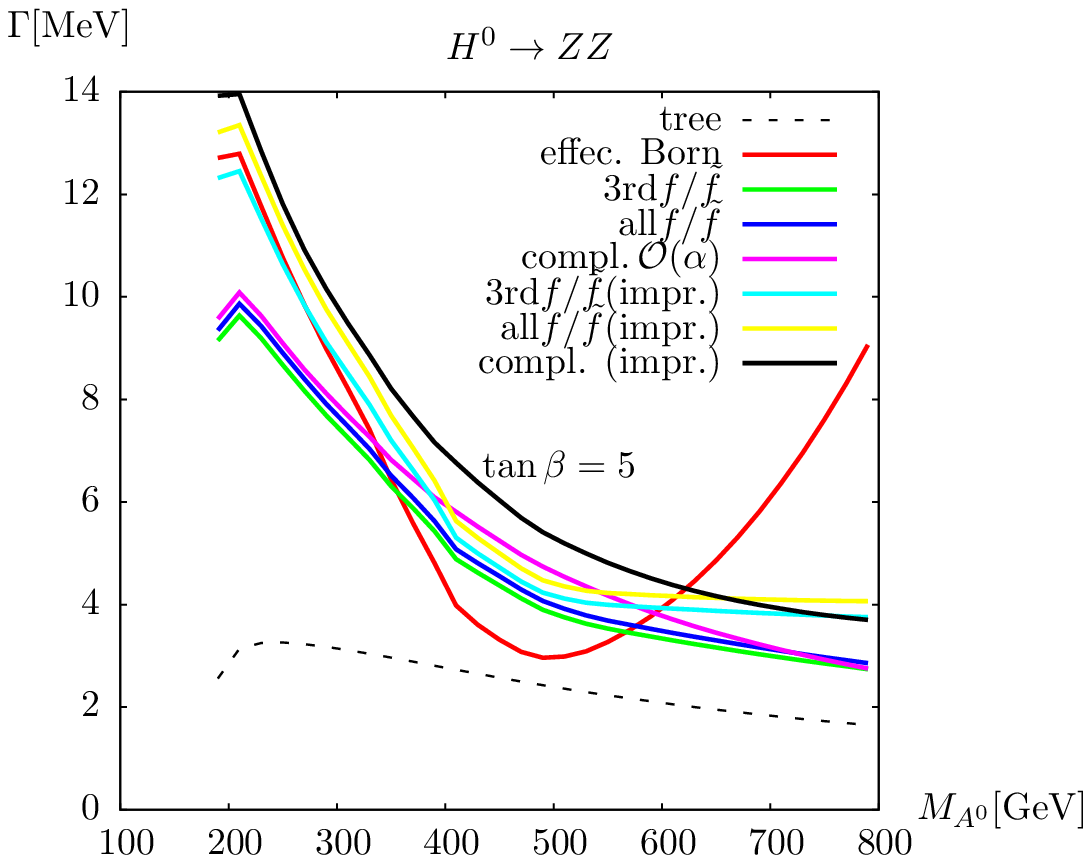}
\hspace{2em}
\includegraphics[width=0.47\textwidth]{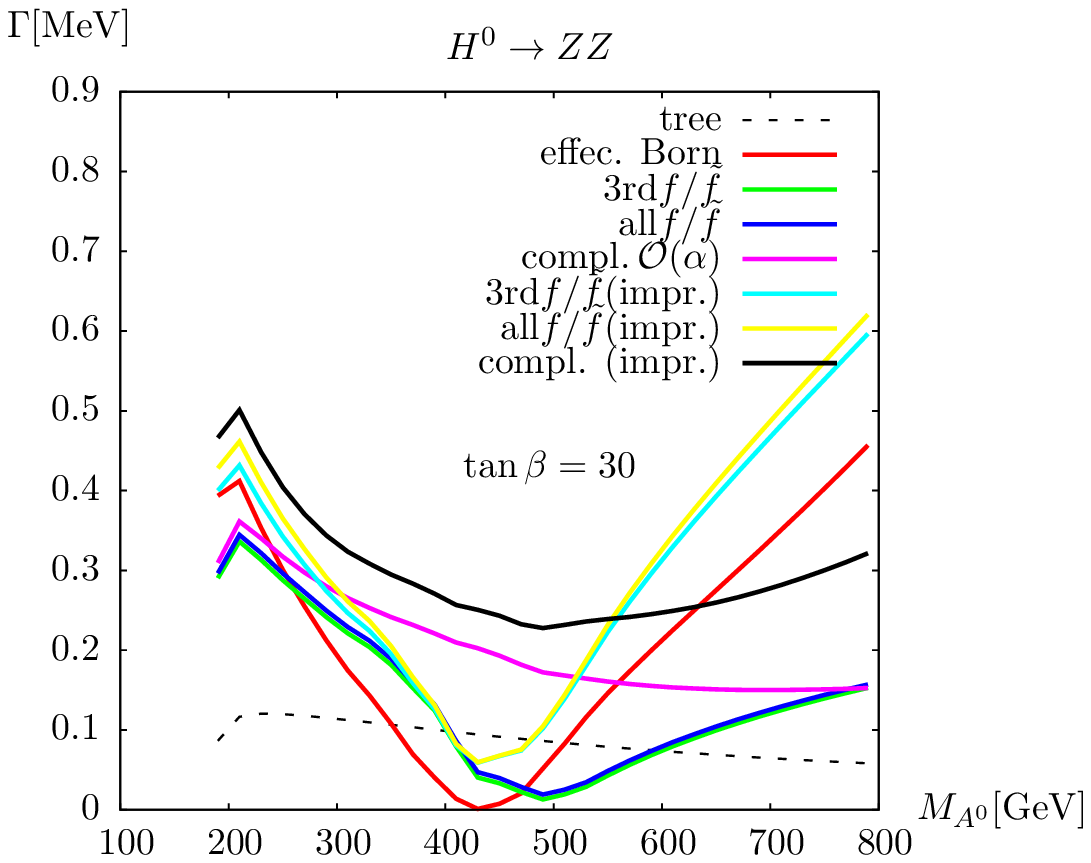}
\caption{The lowest order and the corrected partial decay widths for $H^0\to ZZ$ as a function of $M_{A^0}$ in the $m_h^{\mbox{\small{max}}}\;\mbox{scenario}$ for $\tan\beta=5, 30$. }
\label{HZZwidthmaxcompTB}
\end{figure}

To illustrate the numeric impact of radiative corrections, we show in Fig.~\ref{HWWwidthmaxcompTB} and~\ref{HZZwidthmaxcompTB} the dependence on $M_{A^0}$ of the lowest order and the corrected partial decay widths in the $m_h^{\mbox{\small{max}}}$ scenario for $\tan\beta=5,\;30$. As can be seen by comparing the tree-level and the effective Born results in the figures, the contributions of Higgs propagator corrections are significant, they change the dependence of the tree-level result on $M_{A^0}$ dramatically. An interesting feature of the effective Born results is that they reach a minimum for moderate $M_{A^0}$ values ($\sim420/500$ GeV depending on $\tan\beta$). This is due to the cancellation of the two parts in the effective Born amplitude Eq.~(\ref{oseffbornamp}) at such values of $M_{A^0}$. As shown in the figures, the tree-level partial decay width for $\tan\beta=30$ is much smaller than that for $\tan\beta=5$, since for relatively large $M_{A^0}$, the tree-level coupling of $H^0$ to vector boson pair is suppressed by $\tan\beta$ as well, as explained in the discussions below Eq.~(\ref{HWWtreecoupl}). For $H^0\to WW$, at $\tan\beta=5$ the contribution of the fermionic and sfermionic sector yields the dominant part of the $\mc O(\alpha)$ corrections, and the leading contribution from the fermionic and sfermionic sector is that from the third generation fermions and sfermions. For $\tan\beta=30$, besides the fermionic and sfermionic contribution, the contribution from other sectors to $\mc O(\alpha)$ corrections are also important. In Fig.~\ref{HWWwidthmaxcompTB} we also show the effective one-loop results with fermionic and sfermionic loop contributions (corrections beyond $\mc O(\alpha)$ described in Sec.~\ref{HVVonshell} are included), from which one can see again that the third generation fermionic and sfermionic contribution yields the leading contribution of the fermionic and sfermionic sector. For $\tan\beta=5$, the relative difference between the effective one-loop results and the $\mc O(\alpha)$ ones is less than $50\%$ in the plotted region. For $\tan\beta=30$, the relative difference between the effective one-loop results and the $\mc O(\alpha)$ ones is less than $50\%$ only for small to moderate $M_{A^0}$ values, it can go beyond $100\%$ for large $M_{A^0}$ values and reaches $\sim200\%$ for $M_{A^0}=800$ GeV, as a consequence of the strong suppression of the tree-level coupling by both $\tan\beta$ and $M_{A^0}$. 

Fig.~\ref{HZZwidthmaxcompTB} depicts the partial decay width for $H^0\to ZZ$. The decay width falls off rapidly when $M_{A^0}$ goes below $\sim 200\,\mbox{GeV}$, this is because for such $M_{A^0}$ values, the Higgs boson mass is just above the production threshold of the $Z$ boson bosons, hence the result is strongly suppressed by the available phase space. As in the decay of $H^0\to WW$, for $\tan\beta=5$ the fermionic and sfermionic contribution comprises the dominant part of the $\mc O(\alpha)$ corrections, while for $\tan\beta=30$ the contribution from other sectors becomes important. The relative difference between the effective one-loop results and the $\mc O(\alpha)$ ones is less than $50\%$ in the plotted region for $\tan\beta=5$. For $\tan\beta=30$ the effective one-loop results including the fermionic and sfermionic loop contribution differ significantly from the corresponding $\mc O(\alpha)$ ones at large $M_{A^0}$ values, but the difference between the complete effective one-loop result and the complete $\mc O(\alpha)$ result is smaller, with a relative size less than $50\%$ when $M_{A^0}\lesssim 550$ GeV (it reaches $100\%$ for $M_{A^0}\sim 800$ GeV). Fig.~\ref{HWWwidthmaxcompTB} and~\ref{HZZwidthmaxcompTB} show that although the effective Born decay width can be rather small for certain $M_{A^0}$ values (especially for large $\tan\beta$), the one-loop corrected widths for different values of $M_{A^0}$ turn out to be of comparable size.

\begin{figure}[htbp]
\includegraphics[width=0.47\textwidth]{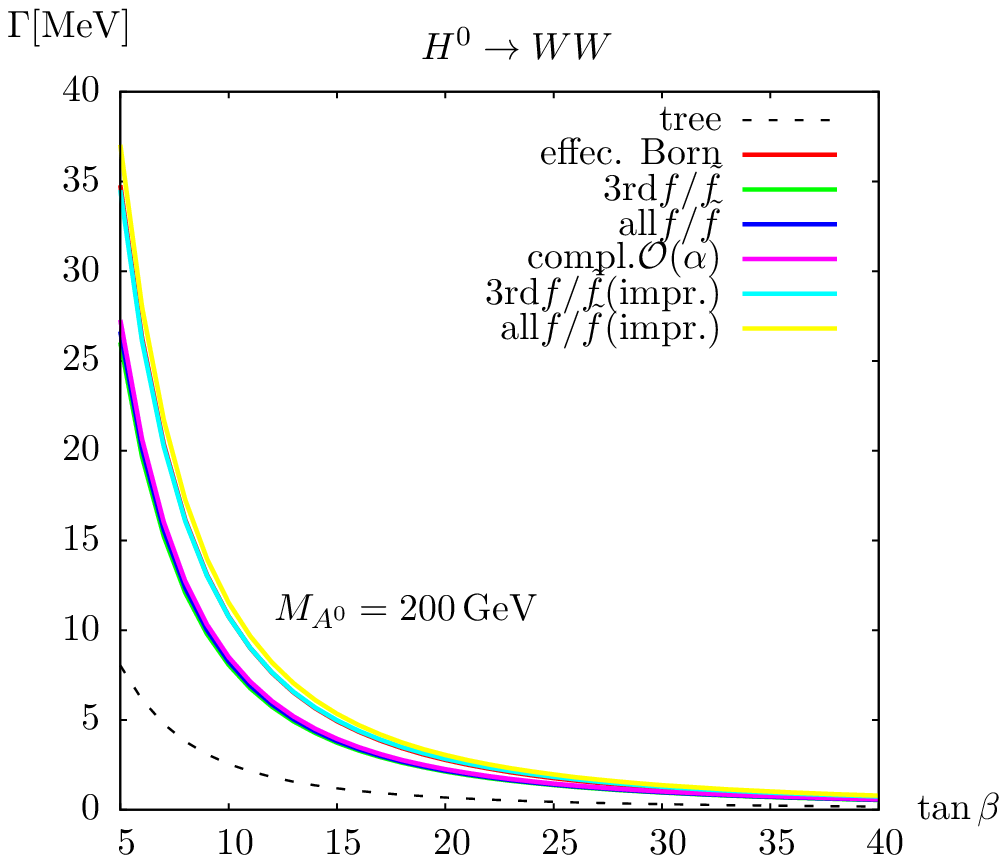}
\hspace{2em}
\includegraphics[width=0.47\textwidth]{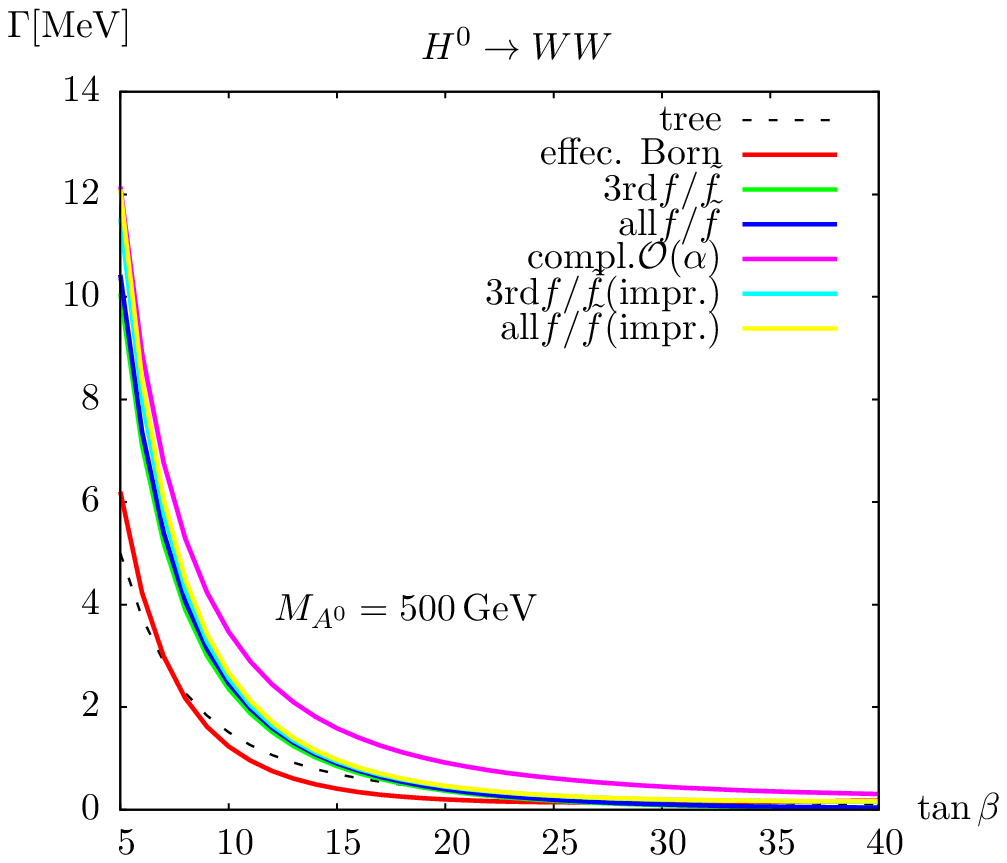}
\caption{The lowest order and the corrected partial decay widths for $H^0\to WW$ as a function of $\tan\beta$ in the $m_h^{\mbox{\small{max}}}\;\mbox{scenario}$ for $M_{A^0}=200, 500\,\mbox{GeV}$.}
\label{HWWwidthmaxcompMA0}
\end{figure}

\begin{figure}[htbp]
\includegraphics[width=0.47\textwidth]{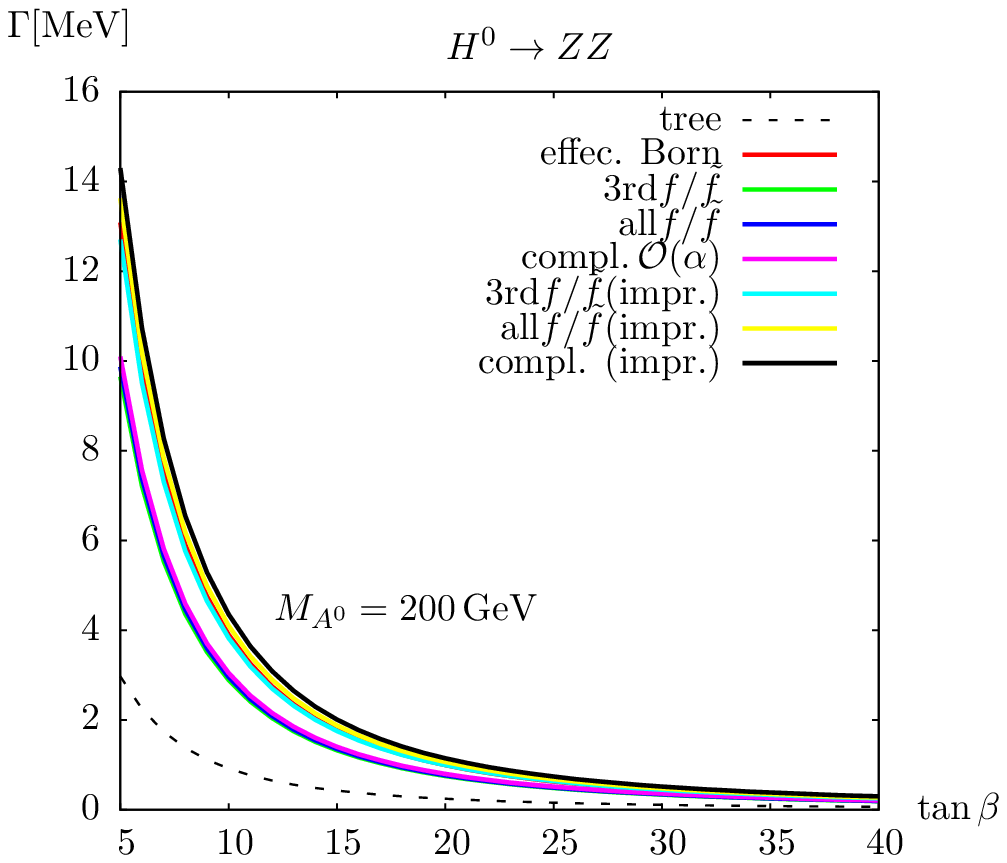}
\hspace{2em}
\includegraphics[width=0.47\textwidth]{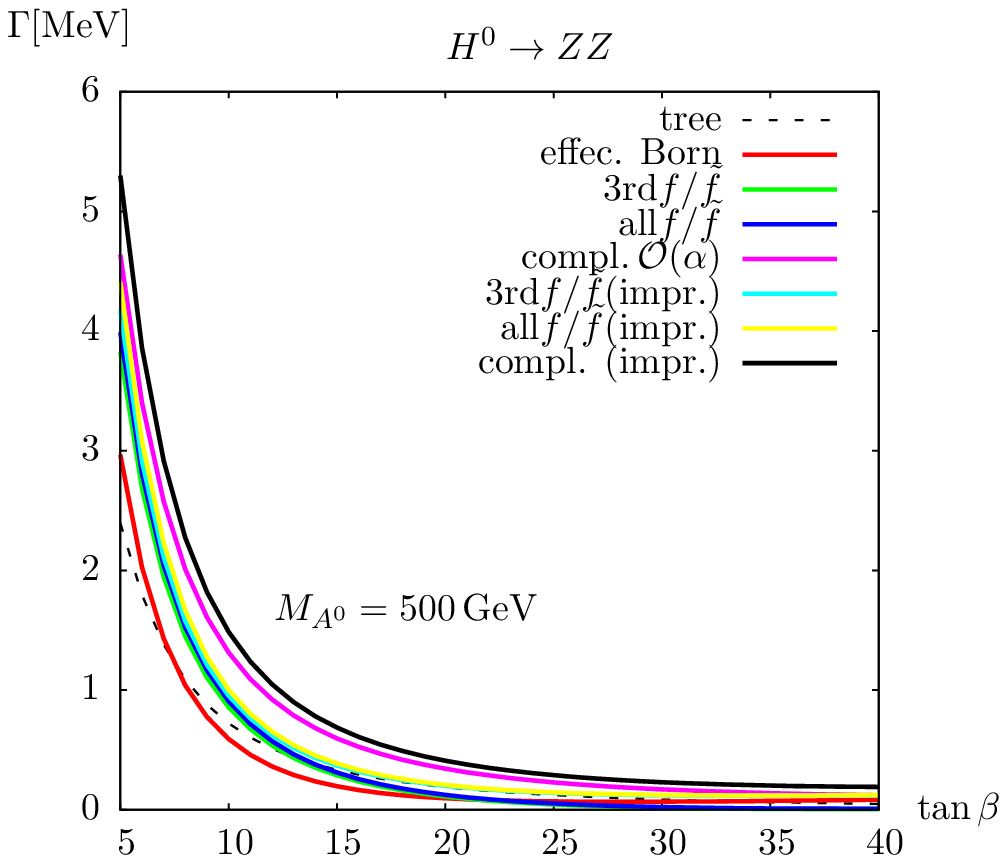}
\caption{The lowest order and the corrected partial decay widths for $H^0\to ZZ$ as a function of $\tan\beta$ in the $m_h^{\mbox{\small{max}}}\;\mbox{scenario}$ for $M_{A^0}=200, 500\,\mbox{GeV}$.}
\label{HZZwidthmaxcompMA0}
\end{figure}

Fig.~\ref{HWWwidthmaxcompMA0} and \ref{HZZwidthmaxcompMA0} show the lowest order and the corrected partial decay widths of $H^0\to WW$ and $H^0\to ZZ$ as a function of $\tan\beta$ in the $m_h^{\mbox{\small{max}}}$ scenario for $M_{A^0}=200,\;500$\,GeV. For both $M_{A^0}$ values, the partial decay width decreases with $\tan\beta$. As in previous plots, the leading contribution from the fermionic and sfermionic sector is from the third generation fermions and sfermions. For $M_{A^0}=200$ GeV, the fermionic and sfermionic contribution comprises the dominant part of the $\mc O(\alpha)$ corrections, while for $M_{A^0}=500$ GeV the contribution from other sectors also becomes important. The relative difference between the (complete) effective one-loop results and the complete $\mc O(\alpha)$ ones is less than $60\%$ for both $M_{A^0}$ values in the figures. The fact that the partial decay width of $H^0\to ZZ$ is smaller than that of $H^0\to WW$ for the same values of $\tan\beta$ and $M_{A^0}$ is due to the presence of identical particles in the $ZZ$ final state.

\begin{figure}[htbp]
\vskip -4cm
\centering
\hspace*{-3cm}
\includegraphics[width=1.2\textwidth]{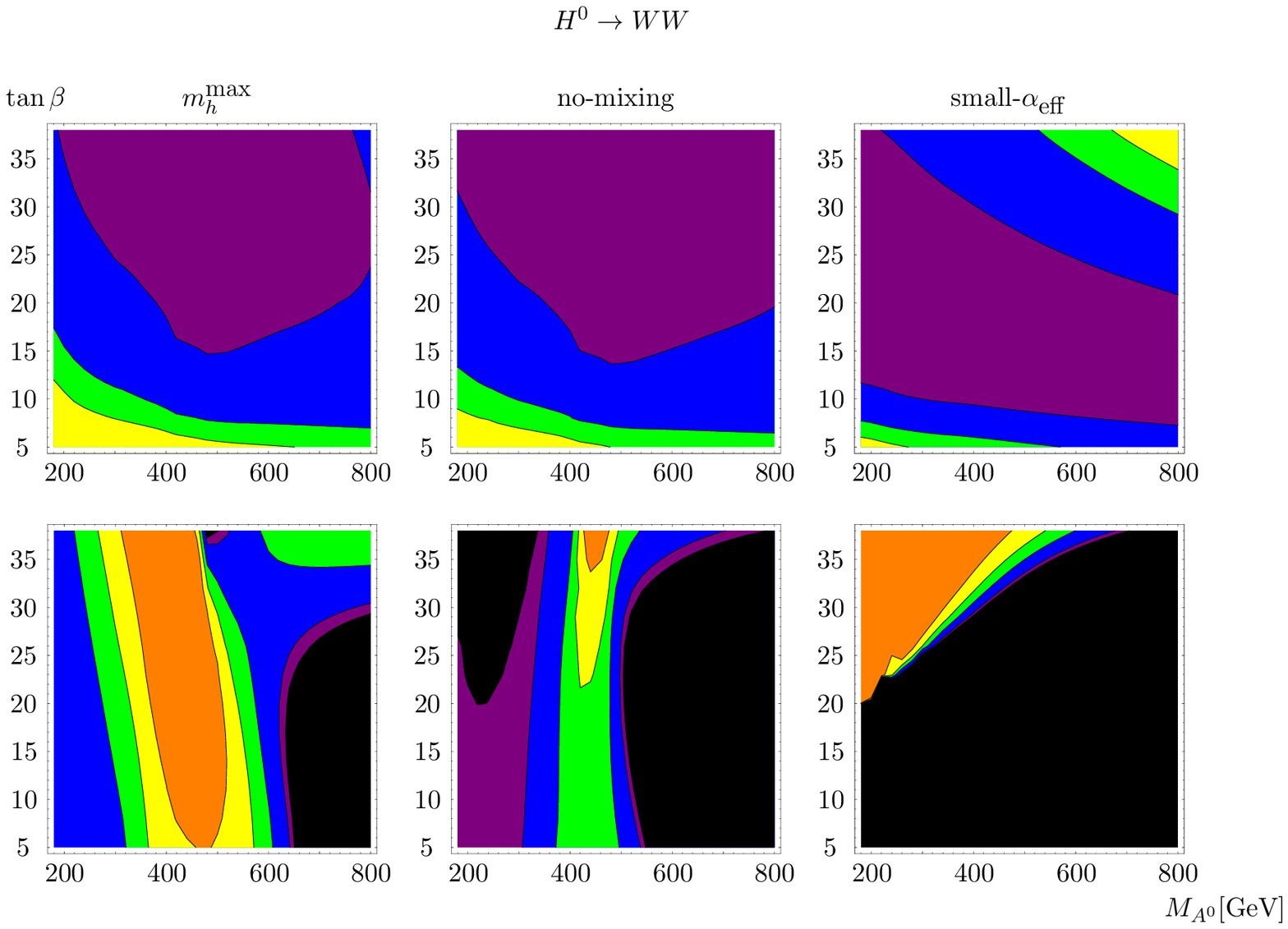}

\caption{Results for the decay width of $H^0\rightarrow WW$ in three scenarios. The upper row shows the corrected decay width (effective one-loop result). The purple region corresponds to $\Gamma_{H^0}<1\mbox{MeV}$, the blue region to $1\mbox{MeV}<\Gamma_{H^0}<5\mbox{MeV}$, the green region to $5\mbox{MeV}<\Gamma_{H^0}<10\mbox{MeV}$, and the yellow region to $10\mbox{MeV}<\Gamma_{H^0}<50\mbox{MeV}$. The lower row shows the corresponding relative correction $\delta$ (divided by the effective Born result). The purple region corresponds to $0<\delta<5\%$, the blue region to $5\%<\delta<25\%$, the green region to $25\%<\delta<50\%$, the yellow region to $50\%<\delta<100\%$,  and the orange region to $\delta>100\%$, the black region corresponds to negative relative correction.}
\label{HWW}
\end{figure}

\begin{figure}[htbp]
\vskip -4cm
\centering
\hspace*{-3cm}
\includegraphics[width=1.2\textwidth]{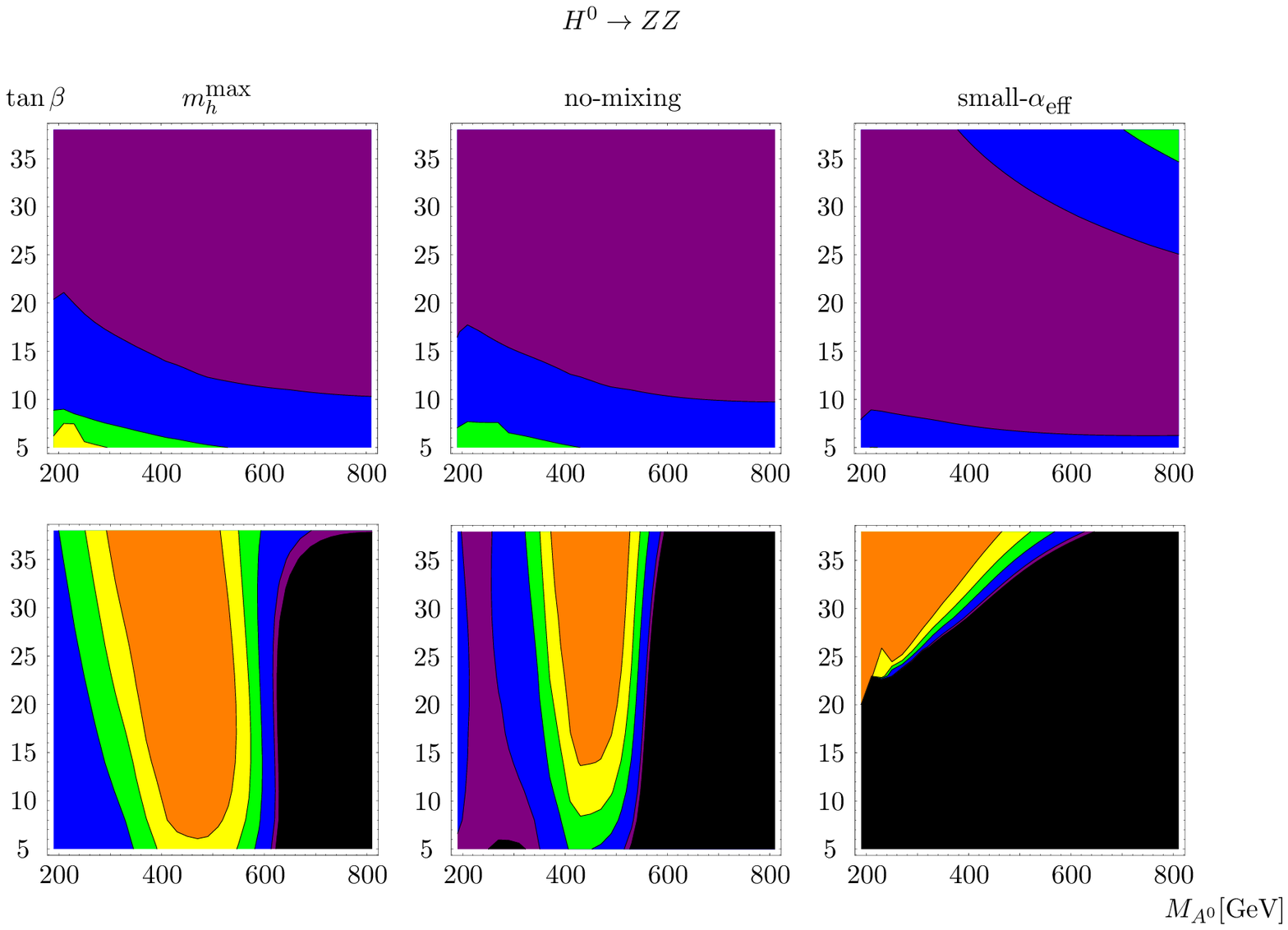}

\caption{Results for the decay width of $H^0\rightarrow ZZ$ in three scenarios. The upper row shows the corrected decay width (effective one-loop results).  The purple region corresponds to $\Gamma_{H^0}<1\mbox{MeV}$, the blue region to $1\mbox{MeV}<\Gamma_{H^0}<5\mbox{MeV}$, the green region to $5\mbox{MeV}<\Gamma_{H^0}<10\mbox{MeV}$, and the yellow region to $10\mbox{MeV}<\Gamma_{H^0}<50\mbox{MeV}$. The lower row shows the corresponding relative correction $\delta$ (divided by the effective Born result). The purple region corresponds to $0<\delta<5\%$, the blue region to $5\%<\delta<25\%$, the green region to $25\%<\delta<50\%$, the yellow region to $50\%<\delta<100\%$,  and the orange region to $\delta>100\%$, the black region corresponds to negative relative correction.}
\label{HZZ}
\end{figure}

In Fig.~\ref{HWW} we show the corrected partial decay width as well as the relative size of the radiative corrections for $H^0\to WW$ in the $M_{A^0}$-$\tan\beta$ plane for three different scenarios. For the size of the width and the relative corrections see the caption of the figure. As illustrated there, in the $m_h^{\mbox{\small{max}}}$ scenario the width is rather small for large $\tan\beta$ in a wide range of $M_{A^0}$ values. It increases when $\tan\beta$ decreases, the relative size of the loop corrections increases rapidly with $M_{A^0}$ and exceeds the effective tree-level result when $M_{A^0}>310\sim460\,\mbox{GeV}$ depending on the values of $\tan\beta$. For large values of $M_{A^0}$ ($\tan\beta\lesssim30$), the relative size of the loop corrections decreases with $M_{A^0}$ and becomes negative when $M_{A^0}\sim 650\,\mbox{GeV}$. In the no-mixing scenario, the corrected width also increases when $\tan\beta$ decreases, and the fermionic and sfermionic contributions are negative over a large fraction of the scanned parameter space. In the small-$\alpha_{\mbox{\small{eff}}}$ scenario, the corrected decay width increases when $\tan\beta$ decreases unless when both $M_{A^0}$ and $\tan\beta$ are large, where the partial decay width is significantly increased by the Higgs propagator corrections. The relative correction is negative in the scanned $M_{A^0}$-$\tan\beta$ plane except in the upper-left corner. At large $\tan\beta$ values, the relative size of the loop corrections increases with $\tan\beta$ and exceeds 100\% quite rapidly. Fig.~\ref{HZZ} illustrates the results for $H^0\to ZZ$ in three different scenarios. The results shown in these figures exhibit similar features to those shown in the plots for $H^0\to WW$, but the corresponding width is smaller due to the presence of identical particles in the final state.

\section{Conclusions}
\label{conclusion}
We have computed the electroweak $\mc O(\alpha)$ radiative corrections to the decay of the heavy CP-even MSSM Higgs boson to weak gauge bosons. Due to the suppression of the tree-level $H^0WW/H^0ZZ$ coupling, the electroweak contributions to the partial decay width are significant, they can easily exceed the tree-level result in certain parameter space. Although the effective Born decay width can be rather small for certain $M_{A^0}$ values (especially for large $\tan\beta$), the corrected partial widths for different values of $M_{A^0}$ are of comparable size. We also presented the effective one-loop results for the partial decay widths, which include corrections beyond $\mc O(\alpha)$. The numeric impact of such corrections is significant for large values of $\tan\beta$ and $M_{A^0}$, while it is less important for small values of $\tan\beta$, and also for large $\tan\beta$ with small to moderate $M_{A^0}$ values.

\bibliographystyle{JHEP}
\bibliography{references}

\newpage

\end{document}